\documentclass[aps,prb,showpacs,twocolumn,amsmath]{revtex4}
\usepackage{latexsym}
\usepackage{amssymb}
\usepackage{graphicx}
\usepackage{bm}

\newcommand{\pdag}{{\phantom{\dagger}}}
\newcommand{\bq}{\begin{equation}}
\newcommand{\eq}{\end{equation}}
\newcommand{\bn}{\begin{eqnarray}}
\newcommand{\en}{\end{eqnarray}}

\begin{document}

\title{Theoretical investigations for shot noise in correlated resonant tunneling 
through a quantum coupled system}

\author{Ivana Djuric$^{1}$}
\author{Bing Dong$^{1,2}$}
\email[Email: ]{bdong@stevens.edu}
\author{H. L. Cui$^{1,3}$}
\affiliation{$^{1}$Department of Physics and Engineering Physics, Stevens Institute of 
Technology, Hoboken, New Jersey 07030 \\
$^{2}$Department of Physics, Shanghai Jiaotong University,
1954 Huashan Road, Shanghai 200030, China \\
$^{3}$School of Optoelectronics Information Science and Technology, Yantai University, 
Yantai, Shandong, China}

\begin{abstract}

In this paper, we carry out a theoretical analysis of the zero-frequency and 
finite-frequency shot noise in electron tunneling through a two-level interacting system 
connected to two leads, when a coherent coupling between the two levels is present, by 
means of recently developed bias-voltage and temperature dependent quantum rate 
equations. 
For this purpose, we generalize the traditional generation-recombination approach for 
shot noise of two-terminal tunneling devices properly to take into account the coherent 
superposition of different electronic states (quantum effects). As applications, 
analytical and numerical investigations have been given in detail for two cases: (1) 
electron tunneling through a quantum dot connected to ferromagnetic leads with intradot 
spin-flip scattering, and (2) spinless fermions tunneling through seriesly coupled 
quantum dots, focusing on the shot noise as functions of bias-voltage and frequency. 

\end{abstract}

\maketitle

\section{INTRODUCTION}

In recent years the study of shot noise in mesoscopic quantum systems has become an 
emerging topic in mesoscopic physics, because measurement of shot noise can reveal more 
information of transport properties which are not available through the conductance 
measurement alone.\cite{Blanter,Beenakker} For example, dynamic correlations in the 
tunneling current originated from the Pauli exclusion principle can provide information 
regarding the barrier geometry. On the other hand, strong Coulomb blockade effect also 
acts in correlating wavepackets and shot noise, when the charging energy becomes larger 
than the thermal energy in small quantum devices. 

Besides, special interest has recently been put upon investigation of tunneling through 
two-level system (TLS), when the coherent coupling between the two levels is presented. 
In such internally coupled system, quantum interference effects between the two levels 
play a crucial role in its transport, inducing, for instance, temporal damped 
oscillations in average current and a corresponding fine structure in shot noise power 
spectrum $S(\omega)$ due to its intrinsic Rabi 
oscillation.\cite{Wiel,Gurvitz,Sun,Brahim,Fedichkin,Gurvitz2,Aguado,Engel} Furthermore, 
it is predicted that a dip or a peak shown in $S(\omega)$ at the Rabi frequency reflects 
the differing relative phase of states carrying the tunneling current.\cite{GurvitzIEEE} 

In literature, the quantum rate equations have been developed to describe this kind of 
quantum coherence effects in quantum tunneling through TLS, in associated with the 
strong Coulomb blockade effect.\cite{Nazarov,Gurvitz1} So far, most calculations for 
shot noise in TLS have been based on the number-resolved version of quantum rate 
equations, which takes numbers of tunneled electrons through TLS as parameters of 
density matrix elements (the degree of freedom of reservoirs), and have made use of 
Laplace transform to calculate the current-current correlation functions. Unfortunately, 
this scheme is only valid at zero temperature and high bias-voltage 
condition.\cite{Brahim,Fedichkin,Gurvitz2,GurvitzIEEE} Therefore, it is quite desirable 
to find a way of evaluating current correlation in arbitrary bias-voltages, particularly 
in moderately small bias-voltage region, where the coherence plays a more prominent role 
in quantum tunneling processes. Accordingly, the purpose of this paper is to develop a 
general scheme to study quantum shot noise spectrum of coupled small quantum systems 
based on the recently established quantum rate equations with a bias-voltage and 
temperature dependent version.\cite{Dong}       

Actually, a numerical method, named generation-recombination approach, was established 
for analysis of shot noise in two-terminal single-electron tunneling devices with the 
classical rate equation (classical shot noise) in Refs. 
[\onlinecite{Hershfield,Korotkov,Hanke}]. These earlier papers proposed a powerful 
method to evaluate the double-time current correlation function by switching the 
time-dependent current to a time evolution propagator of the density matrices in a 
consistent way. In the present paper we will modify this traditional approach to suit 
the underlying quantum rate equations for properly accounting for the quantum coherence 
effects, i.e., the nondiagonal elements of density matrix, in calculation of quantum 
shot noise and develop a tractable computation technique in matrix form.   

The rest of this paper is organized as follows. In section II, we study the quantum shot 
noise for an interacting quantum dot (QD) connected to ferromagnetic leads with intradot 
spin-flip scattering. First we review the model Hamiltonian for this internally coupled 
TLS and the underlying quantum rate equations for description of coherent tunneling in 
sequential picture. Subsequently, we elaborate the general formalism for the quantum 
shot noise in matrix form and present a convenient numerical method for the matrix 
calculation in frequency space with the help of the spectral decomposition technique. In 
the final part of this section, we analyze the zero-frequency Fano factor, which 
quantifies correlations of shot noise with respect to its uncorrelated Poissonian value, 
in several limits, and then provide numerical results for the Fano factor as functions 
of bias-voltage and frequency. In section III, we employ the same procedure to evaluate 
the quantum shot noise in two coupled quantum dots in series (CQD). Finally, a brief 
summary is presented in section IV.

\section{Shot noise in a SINGLE QUANTUM DOT with intradot spin-flip scattering}

\subsection{Model and quantum rate equations}

The system that we study is a single QD with an arbitrary intradot Coulomb interaction 
$U$ connected with two ferromagnetic leads. In this paper, we assume that the tunneling 
coupling between the dot and the leads is weak enough to guarantee no Kondo effect and 
that the QD is in the Coulomb blockade regime. This kind of spin-related single-electron 
devices suffers inevitably intrinsic relaxations (decoherence) due to the spin-orbital 
interaction\cite{Fedichkin} or the hyperfine-mediated spin-flip 
transition.\cite{Erlingsson} For simplicity, we model the intrinsic spin relaxation with 
a phenomenological spin-flip term $R$ and assume that this spin-flip process happens 
solely within the dot (the spin is conserved during tunneling).

Moreover, we assume that the temperature is low enough for one to see the effects due to 
the discrete charging and discrete structure of the energy levels, i.e. $k_{B}T\ll U$, 
$\Delta$ ($\Delta$ is the energy spacing between orbital levels). Each of the two leads 
is separately in thermal equilibrium with the chemical potential $\mu_{\eta}$, which is 
assumed to be zero in the equilibrium condition and is taken as the energy reference 
throughout the paper. In the nonequilibrium case, the chemical potentials of the leads 
differ by the applied bias-voltage $V$. We are interested in the region $eV\ll \Delta$, 
where only one dot level $(\epsilon_{d})$ contribute to the transport. Here we neglect 
Zeeman splitting of the level due to weak magnetic fields $B$ ($g\mu_{B}B<k_{B}T)$, 
which means that both the spin-up and spin-down currents through the dot go through the 
same orbital level $\epsilon_{d}$. Therefore, the Hamiltonian of resonant tunneling 
through a single QD can be written as:\cite{Dong}
\bn H &=& \sum_{\eta,
k, \sigma}\epsilon _{\eta k\sigma} c_{\eta k\sigma }^{\dagger
}c_{\eta k\sigma }^{\pdag}+ \epsilon_{d} \sum_{\sigma}
c_{d \sigma }^{\dagger }c_{d \sigma }^{\pdag}+ R( c_{d \uparrow}^{\dagger} 
c_{d\downarrow}^{\pdag} \cr
&& + c_{d\downarrow}^{\dagger} c_{d\uparrow}^{\pdag}) + Un_{d \uparrow }n_{d \downarrow 
}+\sum_{\eta, k, \sigma} (V_{\eta \sigma} c_{\eta k\sigma }^{\dagger }c_{d 
\sigma}^{\pdag} +{\rm {H.c.}}), \cr && \label{hamiltonian1} 
\en
where $c_{\eta k \sigma}^{\dagger}$ ($c_{\eta k \sigma }$) and $c_{d \sigma}^{\dagger}$ 
($c_{d \sigma}$) are the creation (annihilation) operators for electrons with momentum 
$k$, spin-$\sigma$ and energy $\epsilon_{\eta k \sigma}$ in the lead $\eta$ ($\eta={\rm 
L,R}$) and for a spin-$\sigma$ electron on the QD, respectively. $n_{d\sigma}=c_{d 
\sigma}^{\dagger} c_{d \sigma}^{\pdag}$ is the occupation operator in the QD. The fourth 
term describes the Coulomb interaction among electrons on the QD. The fifth term 
represents the tunneling coupling between the QD and the reservoirs. We assume that the 
coupling strength $V_{\eta \sigma}$ is spin-dependent to take into account the 
ferromagnetic leads.

Under the assumption of weak coupling between the QD and the
leads, and with the application of the wide band limit in the two leads,
electronic
transport through this system in sequential regime can be described by the bias-voltage 
and temperature dependent quantum rate equations for the dynamical evolution of the 
density matrix elements:\cite{Dong}
\begin{subequations}
\label{rateqSQD} \bq
\dot{\rho}_{00}= \sum_{\eta\sigma} ( \Gamma_{\eta\sigma}^{-} \rho_{\sigma \sigma} - 
\Gamma_{\eta\sigma}^{+} \rho_{00}), \label{r0} \eq \bn
\dot{\rho}_{\sigma \sigma} &=& \sum_{\eta}\Gamma_{\eta\sigma}^{+} 
\rho_{00}+\sum_{\eta}\widetilde {\Gamma}_{\bar{\eta\sigma}}^{-} \rho_{dd}- \sum_{\eta}( 
\Gamma_{\eta\sigma}^{-} + \widetilde {\Gamma}_{\eta\bar{\sigma}}^{+}) \rho_{\sigma 
\sigma}\nonumber \\
&& + i R(\rho_{\sigma \bar{\sigma}}-\rho_{\bar{\sigma}\sigma}),
\label{r1} \en \bq
\dot{\rho}_{\sigma \bar{\sigma}} = iR(\rho_{\sigma \sigma} - \rho_{\bar{\sigma} 
\bar{\sigma}})  - \frac{1}{2}\sum_{\eta} ( \widetilde\Gamma_{\eta\sigma}^{+} + 
\widetilde\Gamma_{\eta\bar{\sigma}}^{+} + \Gamma_{\eta\sigma}^{-} + 
\Gamma_{\eta\bar{\sigma}}^{-}) \rho_{\sigma \bar{\sigma}},
\label{r2} \eq \bq
\dot{\rho}_{dd}= \sum_{\eta}\widetilde{\Gamma}_{\eta\downarrow}^{+} \rho_{\uparrow 
\uparrow} + \sum_{\eta}\widetilde {\Gamma}_{\eta\uparrow}^{+} \rho_{\downarrow 
\downarrow} - \sum_{\eta}( \widetilde{\Gamma}_{\eta\uparrow}^{-} + 
\widetilde{\Gamma}_{\eta\downarrow}^{-}) \rho_{dd}, \label{r3} \eq
\end{subequations}
($\sigma=\uparrow,\downarrow$ stands for electron spin and $\bar \sigma$ is the spin 
opposite to $\sigma$). The statistical expectations of the diagonal elements of the 
density matrix, $\rho_{ii}$ ($i=\{0, \sigma, d\}$), give the occupation probabilities of 
the resonant level in the QD as follows: $\rho_{00}$ denotes the occupation probability 
that central region is empty, $\rho_{\sigma\sigma}$ means that the QD is singly occupied 
by a spin-$\sigma$ electron, and $\rho_{dd}$ stands for the double occupation by two 
electrons with different spins. Note that they must satisfy the normalization relation 
$\rho_{00}+ \rho_{dd}+ \sum_{\sigma} \rho_{\sigma \sigma}=1$. The non-diagonal elements 
$\rho_{\sigma \bar\sigma}$ describe the coherent superposition of different spin states. 
These temperature-dependent tunneling rates are defined as $\Gamma_{\eta\sigma}^{\pm}= 
\Gamma_{\eta \sigma} f_{\eta}^{\pm}(\epsilon_{d})$ and 
$\widetilde{\Gamma}_{\eta\sigma}^{\pm}= \Gamma_{\eta \sigma} 
f_{\eta}^{\pm}(\epsilon_{d}+U)$, where $\Gamma_{\eta\sigma}$ are the tunneling 
constants, $f_{\eta}^{+}(\omega)=\{1+e^{(\omega-\mu_{\eta})/T} \}^{-1}$ is the Fermi 
distribution function of the $\eta$ lead and 
$f_{\eta}^{-}(\omega)=1-f_{\eta}^{+}(\omega)$. Here, $\Gamma_{\eta\sigma}^{+}$ 
($\Gamma_{\eta\sigma}^{-}$) describes the tunneling rate of electrons with spin-$\sigma$ 
into (out of) the QD from (into) the $\eta$ lead without the occupation of the 
$\bar{\sigma}$ state. Similarly, $\widetilde {\Gamma}_{\eta\sigma}^{+}$ ($\widetilde 
{\Gamma}_{\eta\sigma}^{-}$) describes the tunneling rate of electrons with spin-$\sigma$ 
into (out of) the QD, when the QD is already occupied by an electron with 
spin-$\bar{\sigma}$, exhibiting the modification of the corresponding rates due to the 
Coulomb repulsion. The particle current $I_{\eta}$ flowing from
the lead $\eta$ to the QD is \bq
I_{\eta}/e=\sum_{\sigma} ( \widetilde{\Gamma}_{\eta \sigma}^{-} \rho_{dd} + \Gamma_{\eta 
\sigma}^{-} \rho_{\sigma \sigma} - \widetilde{\Gamma}_{\eta \bar{\sigma}}^{+} 
\rho_{\sigma \sigma} - \Gamma_{\eta \sigma}^{+} \rho_{00}).
\label{iii} \eq
This formula demonstrates that the current is primarily determined by the diagonal 
elements of the density matrix of the central region. However, the nondiagonal element 
of the density matrix is coupled with the diagonal elements in the rate 
equation~(\ref{r1}), and therefore indirectly influences the
tunneling current.

\subsection{Quantum shot noise formula}

There is a well-established procedure, namely, the generation-recombination approach for 
multielectron channels, for the calculation of the noise power spectrum based on the 
classical rate equations (classical shot noise).\cite{Hershfield,Korotkov,Hanke} In this 
section we modify this approach in order to take into account the nondiagonal density 
matrix elements and derive the general expression for a quantum shot noise for the 
single QD.

It is well known that the noise power spectra can be expressed as the Fourier transform 
of the current-current correlation function: 
\bn
\label{powerspectrum}
S_{I_{\eta}I_{\eta'}}(\omega)&=&2\int_{-\infty}^{\infty}dt e^{i
\omega t}[\langle
I_{\eta}(t)I_{\eta'}(0)\rangle - \langle I_{\eta} \rangle \langle I_{\eta'} \rangle ] 
\nonumber\\
&=& 2\langle I_{\eta}(t) I_{\eta'}(0) \rangle_{\omega}- 2\langle
I_{\eta} \rangle_{\omega} \langle I_{\eta'} \rangle_{\omega}. 
\en
Here, $I_{\eta}$ and $I_{\eta'}$ are the electrical currents
across the $\eta$ and $\eta'$ junctions and $t$ is the time.

Before proceeding with investigation of current correlation, it is helpful to rewrite 
the quantum rate equations as matrix form: 
\bq
\frac{d {\bm \rho}(t)}{dt}= \mathbf{M}\ {\bm \rho}(t)\ ,
 \label{rate}
\eq
where ${\bm \rho}(t)=(\rho_{00}, \rho_{\uparrow \uparrow}, \rho_{\downarrow \downarrow}, 
\rho_{dd}, \rho_{\uparrow \downarrow}, \rho_{\downarrow \uparrow})^{T}$ is a vector 
whose components are the density matrix elements, and the $6 \times 6$ matrix ${\bf M}$ 
can be easily obtained from Eqs.~(\ref{rateqSQD}). Therefore, the statistical averaging 
of an any time-dependent operator $\hat{A}(t)$ should be replaced by the summation over 
all elements in this new ${\bm \rho}(t)$ representation: 
\bq
\langle\hat{A}(t)\rangle = {\rm Tr} \{\hat{A} \rho \} = \sum_{k} [{\bm A} {\bm \rho}(t) 
]_{k}, 
\eq
in which ${\bm A}$ is the matrix expression of the operator $\hat {A}$ under the basis 
$(\rho_{00}, \rho_{\uparrow \uparrow}, \rho_{\downarrow \downarrow}, \rho_{dd}, 
\rho_{\uparrow \downarrow}, \rho_{\downarrow \uparrow})^{T}$, and the summation goes 
over all vector $[{\bm A} {\bm \rho}(t)]$ elements ($k=1,2,\cdots,6$). Correspondingly, 
we can write the average electrical currents across the left ($I_{L}$) and right
($I_{R}$) junctions at time $t$ as: 
\bq
\langle I_{L(R)}(t)\rangle=e \sum_{k} \left[ \hat{\Gamma}_{L(R)} {\bm \rho}(t)
 \right]_{k},
 \label{current}
\eq 
where $\hat{\Gamma}_{L}$ and $\hat{\Gamma}_{R}$ are current
operators and the summation goes over all vector
$[\hat{\Gamma}_{L(R)}{\bm \rho}(t)]$ elements ($k=1,2,\cdots,6$).
The current operators contain the rates for tunneling across the
left and right junctions respectively,\cite{Hershfield} and they
can be read from Eq.~(\ref{iii}) as follows:
$\tilde{\Gamma}_{\eta\sigma}^{-}$ describes the tunneling rate of
an electron with spin-$\sigma$ out of the QD, when the QD were
previously occupied by two electrons. After this tunneling event,
QD become singly occupied by a spin-$\bar{\sigma}$ electron. In
the other words, $\tilde{\Gamma}_{\eta\sigma}^{-}$ is a transition
rate from $\rho_{dd}$ to the $\rho_{\bar{\sigma}\bar{\sigma}}$. It
increases occupation probability $\rho_{\bar{\sigma}\bar{\sigma}}$
[this can be also seen by looking the second term of
Eq.~(\ref{r1})] and decreases occupation probability $\rho_{dd}$
[the last term in Eq.~(\ref{r3})]. The term
$\Gamma_{\eta\sigma}^{-}\rho_{\sigma\sigma}$ in Eq.~(\ref{iii})
represents the tunneling of $\sigma$ electron out of the QD which
leaves QD empty and $\Gamma_{\eta\sigma}^{-}$ can be interpreted
as a transition rate from $\rho_{\sigma\sigma}$ to the
$\rho_{00}$. In similar way, $\tilde{\Gamma}_{\eta\bar{\sigma}}^{+}$ describes the 
transition
rate from $\rho_{\bar{\sigma}\bar{\sigma}}$ to the $\rho_{dd}$ and
$\Gamma_{\eta\sigma}^{+}$ is a transition rate from $\rho_{00}$ to
the $\rho_{\sigma\sigma}$. In the basis $(\rho_{00}, \rho_{\uparrow \uparrow}, 
\rho_{\downarrow \downarrow}, \rho_{dd}, \rho_{\uparrow \downarrow}, \rho_{\downarrow
\uparrow})^{T}$ , the current operators have a matrix form: 
\bq
\hat{\Gamma}_{\eta}=\pm\left(
\begin{array}{cccccc}
  0 & \Gamma_{\eta\uparrow}^{-} & \Gamma_{\eta\downarrow}^{-} & 0 & 0 & 0 \\
  -\Gamma_{\eta\uparrow}^{+} & 0 & 0 & \tilde{\Gamma}_{\eta\downarrow}^{-} & 0 & 0 \\
  -\Gamma_{\eta\downarrow}^{+} & 0 & 0 & \tilde{\Gamma}_{\eta\uparrow}^{-} & 0 & 0 \\
  0 & -\tilde{\Gamma}_{\eta\downarrow}^{+} & -\tilde{\Gamma}_{\eta\uparrow}^{+} & 0 & 0 
& 0 \\
  0 & 0 & 0 & 0 & 0 & 0 \\
  0 & 0 & 0 & 0 & 0 & 0 \\
\end{array}
\right), 
\label{current operators} 
\eq 
where the sign of the
current is chosen to be positive when the direction of the current
is from left to right, so that the $+$ sign in the last equation
is for $\eta=L$ and the $-$ sign stands for $\eta=R$. The
stationary current can be obtained as: 
\bn 
\label{stationary current}
I = e\sum_{k} \left [\hat{\Gamma}_L {\bm \rho}^{(0)}\right ]_{k} = e \sum_{k} \left [ 
\hat{\Gamma}_R {\bm \rho}^{(0)}\right ]_{k}, 
\en 
where ${\bm \rho}^{(0)}$ is the steady state solution of Eq.~(\ref{rate}) and
which can be obtained from: 
\bn 
\label{steadystate} 
\mathbf{M}{\bm \rho}^{(0)}=0, 
\en
along with the normalization relation $\sum_{n=1}^{4} {\bm \rho}^{(0)}_{n}=1$. We would 
like to point out that in our quantum version of rate equations, it is easy to check 
$\sum_{n} {\bf M}_{nm}=0\ (m=1,2,3,4)$, which implies that: 1) the Matrix ${\bf M}$ has 
a zero eigenvalue; 2) there is always a steady state solution ${\bm \rho}^{(0)}$; 3) the 
normalization relation $\sum_{n=1}^{4} {\bm \rho}_{n}(t)=1$ is
independent on time.

A convenient way to evaluate the double-time correlation function 
Eq~(\ref{powerspectrum}) is to define the propagator $\hat{T}(t)=\exp[{\bf M}t]$, which 
governs the time evolution of the density matrix elements ${\bm \rho}_{k}(t)$. The 
average value of the
electrical currents across the left ($I_{L}$) and the right
($I_{R}$) junctions at a time $t$ is given by: 
\bq
\langle{I_{L(R)}(t)}\rangle=-e \sum_{k} \left[
\hat{\Gamma}_{L(R)}\hat{T}(t) {\bm \rho}^{(0)}\right]_{k},
\label{current1} 
\eq
which allows us to switch the time evolution from the vector ${\bm \rho}(t)$ to the 
current operators. Thus, we identify
$\hat{\Gamma}_{L(R)}(t)=\hat{\Gamma}_{L(R)}\hat{T}(t)$ as the
time-dependent current operators. With these time-dependent
operators we can calculate correlation functions of two current
operators taken at different moments in time. In particular,
correlation function of the currents $I_{\eta}$ and $I_{\eta'}$ in
the tunnel
junctions $\eta$ and $\eta'$, measured at the two times $t$ and $0$ respectively, is 
given by:\cite{Hershfield} 
\bn 
\langle{I_{\eta}(t)I_{\eta'}(0) }\rangle
&=& \theta(t)\sum_{k} [\hat{\Gamma}_{\eta}(t)\hat{\Gamma}_{\eta'} {\bm \rho}^{(0)}]_{k} 
\cr
&+& \theta(-t)\sum_{k}[\hat{\Gamma}_{\eta'}(-t)\hat{\Gamma}_{\eta} {\bm \rho}^{(0)}]_{k} 
\nonumber\\
&=& \theta(t) \sum_{k} [\hat{\Gamma}_{\eta} \hat{T}(t) \hat{\Gamma}_{\eta'} {\bm 
\rho}^{(0)} ]_{k} \cr
&+&  \theta(-t)\sum_{k} [\hat{\Gamma}_{\eta'} \hat{T}(-t) \hat{\Gamma}_{\eta} {\bm 
\rho}^{(0)}]_{k},
\label{current correlation} 
\en 
where
$\theta(t)$ is the Heaviside function and the two terms in
Eq.~(\ref{current correlation}) stand for $t>0$ and for $t<0$. The
Fourier transform of propagator $\hat{T}(\pm t)$ is $\hat{T}(\pm
\omega)=\left(\mp i\omega\hat{I}-\bf{M}\right)^{-1}\ $, where
$\hat{I}$ is an unit matrix. We can further simplify this
expression by using the spectral decomposition of the matrix
$\bf{M}$: 
\bq
\mathbf{M} = \sum_{n} \lambda_{n} \mathbf{S} \mathbf{E}^{(nn)} \mathbf{S}^{-1} = 
\sum_{\lambda} \lambda \hat{P}_{\lambda}, 
\eq 
in which $\lambda$ are
eigenvalues of the matrix $\bf{M}$, $\mathbf{S}$ is a matrix whose
columns are eigenvectors of $\bf{M}$, $\mathbf{E}^{(nn)}$ is a
matrix that has $1$ at $nn$ place and all other elements are
zeros, and $\hat{P}_{\lambda}$ is
a projector operator associated with the eigenvalue $\lambda$, so that $\hat{T}(\pm 
\omega)$ is 
\bq 
\hat{T}(\pm \omega) = \sum_{\lambda}
\frac{\hat{P}_{\lambda}}{\mp i\omega-\lambda}. 
\eq 
Inserting
expression for propagator $\hat{T}$ into Eq.~(\ref{current
correlation}) current-current correlation in the $\omega$-space
becomes 
\bn 
\label{cc} 
\langle I_{\eta}(t) I_{\eta'}(0)
\rangle_{\omega} &=& \sum_{\lambda,k}
\left [ \frac{\hat{\Gamma}_{\eta}\hat{P}_{\lambda}\hat{\Gamma}_{\eta'}} 
{{-i\omega-\lambda}} {\bm \rho}^{(0)} \right ]_{k} \cr
&+& \sum_{\lambda,k} \left [ 
\frac{\hat{\Gamma}_{\eta'}\hat{P}_{\lambda}\hat{\Gamma}_{\eta}} {{i\omega-\lambda}} {\bm 
\rho}^{(0)} \right ]_{k}. 
\en
Eventually, substituting Eq.~(\ref{cc}) into the noise definition 
Eq.~(\ref{powerspectrum}), and noting that in the summation over eigenvalues the 
zero-eigenvalue contribution is canceled exactly by the term $\langle I_{\eta} \rangle 
\langle I_{\eta'} \rangle$, we can obtain the final expression for a noise power 
spectrum: 
\bn
\label{noise} 
&& S_{I_{\eta}I_{\eta'}}(\omega)=\delta_{\eta\eta'}S^{\rm Sch}_{\eta}\nonumber\\
&& + 2\sum_{k,\lambda \neq 0} \left(
\frac{\left [ \hat{\Gamma}_{\eta} \hat{P}_{\lambda}\hat{\Gamma}_{\eta'} {\bm \rho}^{(0)} 
\right ]_{k}}{{-i\omega-\lambda}} + \frac{ \left [ \hat{\Gamma}_{\eta'} 
\hat{P}_{\lambda} \hat{\Gamma}_{\eta} {\bm \rho}^{(0)} \right
]_{k}} {{i\omega-\lambda}}
\right), \nonumber \\
\en
where $S^{\rm Sch}_{\eta}$ is the frequency-independent Schottky noise originating from 
the self-correlation of a given tunneling event with itself, which the double-time 
correlation function Eq.~(\ref{current correlation}) can not contain. Due to the fact 
that the current has no explicit dependence on the nondiagonal elements of the density 
matrix, it can be simply written as:\cite{Hershfield} 
\bn
\label{Schottky}
S^{\rm Sch}_{\eta} = \sum_{k} \left| \left [ \hat{\Gamma}_{\eta} {\bm \rho}^{(0)}  
\right ]_{k}\right|. 
\en

The shot noise power spectrum is given by: 
\bn 
\label{SN}
S(\omega) & =& \alpha^{2} S_{I_{L}I_{L}} (\omega) + \beta^{2}S_{I_{R}I_{R}}(\omega) 
\nonumber \\
&& + \alpha\beta S_{I_{L}I_{R}}(\omega) + \alpha\beta
S_{I_{R}I_{L}}(\omega).
\en 
Here, the coefficients $\alpha$ and
$\beta$, $\alpha+\beta=1$, depend on barriers
geometry.\cite{Davies} For simplicity we take $\alpha=\beta=1/2$.
The Fano factor, which measures a deviation from the uncorrelated
Poissonian noise, is defined as: 
\bn
F(\omega)=\frac{S(\omega)}{2eI}, \label{Fano factor} 
\en 
where
$2eI$ is the Poissonian noise.

\subsection{Discussions and numerical calculations}

In the following we consider two magnetic configurations: the
parallel (P), when the majority of electrons in both leads point
in the same direction, chosen to be the electron spin-up state,
$\sigma=\uparrow$;
and the antiparallel (AP), in which the magnetization of the right electrode is 
reversed. The ferromagnetism of the leads can be accounted for by
means of polarization-dependent coupling constants. Thus, we set
for the P alignment 
\bq
\Gamma_{L\uparrow}=\Gamma_{R\uparrow}=(1+p)\Gamma, \, 
\Gamma_{L\downarrow}=\Gamma_{R\downarrow}=(1-p)\Gamma, \label{pcon} 
\eq 
while
for the AP-configuration we choose 
\bq
\Gamma_{L\uparrow}=\Gamma_{R\downarrow}=(1+p)\Gamma, 
\,\Gamma_{L\downarrow}=\Gamma_{R\uparrow}=(1-p)\Gamma. \label{apcon}
\eq
Here, $\Gamma$ denotes the tunneling coupling between the QD and the leads without any 
internal magnetization, whereas $p$ ($0\leq p< 1$) stands for the polarization strength 
of the leads. We work in the wide band limit, i.e., $\Gamma$ is
supposed to be a constant, and we use it as an energy unit in the
rest of this section.
The zero of energy is chosen to be the Fermi level of the leads in the equilibrium 
condition ($\mu_{L}=\mu_{R}=0$). The
bias-voltage, $V$, between the source and the drain is considered
to be applied symmetrically, $\mu_{L}=-\mu_{R}=eV/2$. The shift of
the discrete level due to the external bias is neglected.

As a reference case for our analysis we use the analytic result
for the case of paramagnetic electrodes, $p=0$. In the system with
paramagnetic electrodes, both channels for electrons with the spin
up ($\uparrow $) and down ($\downarrow$) are equivalent and the
tunneling rates are
$\sum_{\eta} \Gamma_{\uparrow}^{\pm}= \sum_{\eta} \Gamma_{\downarrow}^{\pm} = 
\Gamma^{\pm}$
and $\sum_{\eta} \tilde{\Gamma}_{\uparrow}^{\pm} = \sum_{\eta} 
\tilde{\Gamma}_{\downarrow}^{\pm} = \tilde{\Gamma}^{\pm}$.
We assume that Coulomb repulsion is large and discuss the two
special cases separately. First, we consider that no doubly
occupied state is available in the QD, i.e, $ \rho_{dd}=0$. In
this case Fermi levels $\mu$ of the two leads is bellow
$U+\epsilon_{d}$, meaning $\widetilde{\Gamma}_{\sigma}^{+}\simeq 0$, 
$\widetilde{\Gamma}_{\sigma}^{-} \simeq \sum_{\eta} \Gamma_{\eta \sigma}$. Then the 
quantum rate equations
Eq.~(\ref{rateqSQD}) become:
\begin{subequations}
\label{rateqsingle} 
\bn
\dot{\rho}_{00} &=& \Gamma^{-}{\rho}_{\upuparrows}+\Gamma^{-}{\rho}_{\downdownarrows}-2 
\Gamma^{+}{\rho}_{00}, \label{rs3} \\
\dot{\rho}_{\upuparrows} &=& -\Gamma^{-}{\rho}_{\upuparrows} + \Gamma^{+}{\rho}_{00} - 
2R{\rho}_{\uparrow\downarrow}^{(i)},\label{rs1}\\
\dot{\rho}_{\downdownarrows} &=& -\Gamma^{-}{\rho}_{\downdownarrows} + 
\Gamma^{+}{\rho}_{00} + 2R{\rho}_{\uparrow\downarrow}^{(i)},\label{rs2} \\
\dot{\rho}_{\uparrow\downarrow}^{(r)} &=& -\Gamma^{-}{\rho}_{\uparrow\downarrow}^{(r)}, 
\label{rs4} \\
\dot{\rho}_{\uparrow\downarrow}^{(i)} &=& R(\rho_{\upuparrows}-\rho_{\downdownarrows})- 
\Gamma^{-}{\rho}_{\uparrow\downarrow}^{(i)},\label{rs5}
\en
\end{subequations}
where ${\rho}_{\uparrow\downarrow}^{(r)}$ and
${\rho}_{\uparrow\downarrow}^{(i)}$ stand for real and imaginary
part of non-diagonal density matrix elements and we assume that the
spin-flip term, $R$, is real. The matrix $\textbf{M}$ is given as:
\bq
\mathbf{M}=\left(
\begin{array}{ccccc}
 \Gamma^{-} & \Gamma^{-} & -2\Gamma^{+} & 0 & 0 \\
 -\Gamma^{-} & 0 & \Gamma^{+} & 0 & -2R\\
  0 & -\Gamma^{-} & \Gamma^{+} & 0 & 2R \\
  0 & 0 & 0 & -\Gamma^{-} & 0 \\
  R & -R & 0 & 0 & -\Gamma^{-} \\
\end{array}
\right).
\eq
The eigenvalues of the matrix $\mathbf{M}$ are:
$\lambda_{0}=0$, $\lambda_{1}=-\Gamma^{-}$, $\lambda_{2}=-(\Gamma^{-}+2\Gamma^{+})$, 
$\lambda_{3}=\lambda_{4}^{\ast}=-(\Gamma^{-}+2iR)$
 and the corresponding eigenvectors are given as the columns of
 matrix $\mathbf{S}$:
\bq
\mathbf{S}=\left(
\begin{array}{ccccc}
  \frac{\Gamma^{+}}{[(\Gamma^{-})^{2}+2(\Gamma^{+})^{2}]^{1/2}} & 0 & 
-\frac{1}{\sqrt{6}} & -\frac{i}{\sqrt{3}} & \frac{i}{\sqrt{3}} \\
  \frac{\Gamma^{+}}{[(\Gamma^{-})^{2}+2(\Gamma^{+})^{2}]^{1/2}} & 0 & 
-\frac{1}{\sqrt{6}} & \frac{i}{\sqrt{3}} & -\frac{i}{\sqrt{3}} \\
  \frac{\Gamma^{-}}{[(\Gamma^{-})^{2}+2(\Gamma^{+})^{2}]^{1/2}} & 0 & \frac{2}{\sqrt{6}} 
& 0 & 0 \\
  0 & 1 & 0 & 0 & 0 \\
  0 & 0 & 0 & 1 & 1 \\
\end{array}
\right).
\eq 
The current operators can be read from Eq.~(\ref{iii})
and the nonzero elements are
$(\hat{\Gamma}_{L})_{21}=-{\Gamma}_{L}^{+}$,
$(\hat{\Gamma}_{L})_{12}={\Gamma}_{L}^{-}$,
$(\hat{\Gamma}_{L})_{31}=-{\Gamma}_{L}^{+}$,
$(\hat{\Gamma}_{L})_{13}={\Gamma}_{L}^{-}$,
and
$\hat{\Gamma}_{R}=-\hat{\Gamma}_{L}$.
On the other hand, the steady states can be obtained from
Eq.~(\ref{rateqsingle}), together with
$\rho_{00}+\sum_{\sigma}\rho_{\sigma\sigma}=1$, as:
\begin{subequations}
\label{steadyp} 
\bn
\rho_{\upuparrows}^{0} &=& \rho_{\downdownarrows}^{0} = \frac{\Gamma^{+}}{\Gamma^{-}+ 2 
\Gamma^{+}}, \\
\rho_{00}^{0} &=& \frac{\Gamma^{-}}{\Gamma^{-}+2\Gamma^{+}},
\en
\end{subequations}
and the stationary tunneling current is
$I_{L}=I_{R}=\frac{2e}{\Gamma^{-}+2\Gamma^{+}}(\Gamma_{L}^{-}\Gamma_{R}^{+}-\Gamma_{L}^{
+}\Gamma_{R}^{-})$.
Finally, applying the spectral decomposition of $\mathbf{M}$ and
using Eq.~(\ref{noise}) one gets the Fano factor: 
\bq
\label{fano1}
F=1+\frac{4(\Gamma_{R}^{+}\Gamma_{L}^{-}-\Gamma_{R}^{-}\Gamma_{L}^{+})}{(\Gamma^{-} + 
2\Gamma^{+})^2}+
\frac{2\Gamma_{R}^{+}\Gamma_{L}^{-}}{\Gamma_{R}^{-}\Gamma_{L}^{+}-\Gamma_{R}^{+}\Gamma_{
L}^{-}}.
\eq
In the case of large voltage, i.e., $eV/2\gg\epsilon_{d}$,
$\Gamma_{L}^{-}=\Gamma_{R}^{+}=0$, and $\Gamma_{L}^{+}=\Gamma_{L}$,
$\Gamma_{R}^{-}=\Gamma_{R}$. The current and the Fano factor
are
\bn
\label{current1.1}
I_{1} &=& \frac{\Gamma_{L}\Gamma_{R}}{2\Gamma_{L}+\Gamma_{R}},
\label{fano1.1} \\
F_{1} &=& 1-\frac{4\Gamma_{L}\Gamma_{R}}{(\Gamma_{L}+2\Gamma_{R})^2}.
\en
The Fano factor depends only on the asymmetry in the coupling between
the leads and the dot: it is equal to $\frac{5}{9}$ for the
completely symmetric case $\Gamma_{L}=\Gamma_{R}$, and approaches
$1$ when one of the coupling constants becomes much larger than
the other one.

In the opposite region, when the energies $\epsilon_{d}$ and
 $\epsilon_{d}+U$ are far below the Fermi level $\mu$,
 $eV/2\gg\epsilon_{d}+U$, we have $\Gamma_{L}^{-}=\Gamma_{R}^{+}=0$,
 $\tilde{\Gamma}_{L}^{-}=\tilde{\Gamma}_{R}^{+}=0$, and
 $\tilde{\Gamma}_{L}^{+}=\Gamma_{L}^{+}=\Gamma_{L}$,
 $\tilde{\Gamma}_{R}^{+}=\Gamma_{R}^{+}=\Gamma_{R}$. Substituting
 these expressions  into the rate equations and performing similar
 calculations, the electrical current and the Fano factor are found to be
\bn
I_{2} &=& \frac{\Gamma_{L}\Gamma_{R}}{\Gamma_{L}+\Gamma_{R}},  \label{current1.2} \\
F_{2} &=& 1-2\frac{\Gamma_{L}\Gamma_{R}}{(\Gamma_{L}+\Gamma_{R})^2}.  \label{fano1.2}
\en
The Fano factor is equal to $\frac{1}{2}$ for completely symmetric
coupling and to $1$ for the asymmetric ones.

When the leads are made of paramagnetic materials, the Fano factor
does not depend on the spin-flip process and the same result was
obtained by using the classical rate equations.\cite{Hershfield}
However, this is not true for the ferromagnetic leads.

Now we proceed our investigation of the Fano factor for the system where leads are 
ferromagnetic.
Our numerical calculations for the current-voltage characteristic and the
dependence of the Fano factor on bias-voltage for P- and AP-configurations are presented 
in Figs.~1--3. In these calculations we set
$\epsilon_{d}=1$, Coulomb interaction $U=4$, and temperature $T=0.1$. By increasing the 
bias-voltage between two leads, two steps in
the current-voltage characteristic occur: one is when the Fermi
level of the source, $\mu_{L}$, crosses the discrete levels
$\epsilon_{d}$ (for $eV/2>\epsilon_{d}$) and the other is when the
Fermi level $\mu_{L}$ crosses $\epsilon_{d}+U$ (for
$eV/2>\epsilon_{d}+U$).

The effects of the polarization on the Fano factor without the
spin-flip scattering ($R=0$) are plotted in Fig.~1. An increase of
the polarization will lead to an enhancement of the current noise
in both configurations (P and AP) but for different reasons. Let
us discuss the P- and the AP-configurations separately.

\begin{figure}[htb]
\includegraphics[height=3.in,width=3.5in]{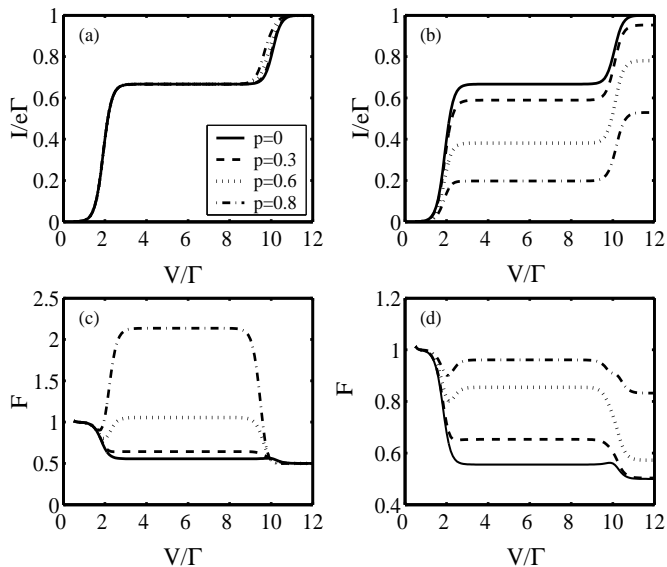}
\caption{Current (a, b) and Fano factor (c, d) vs. bias-voltage in the P-configuration 
(a, c) and AP-configuration (b, d)
calculated for different polarization $p$ without spin-flip process. Other parameters 
are: $\epsilon_{d}=1$, $U=4$, and 
$T=0.1$.}\label{FIG.1}
\end{figure}

When the leads are in the P-configuration [Fig.~1(a) and (c)], an
increase of the polarization will raise the tunneling rates
of electrons with
the spin-up, $\Gamma_{L\uparrow}$ and $\Gamma_{R\uparrow}$, but reduce the tunneling 
rates of spin-down electrons,
$\Gamma_{L\downarrow}$ and $\Gamma_{R\downarrow}$. Accordingly, this will induce an 
increase of the spin-up current and a decrease of the
spin-down current but it will not affect the total current through
the system as shown in Fig.~1(a), because the total current is equal to the summation of 
the spin-up and spin-down currents. In the case that the Coulomb interaction prevents a 
double occupancy of the dot, there will be competition between
tunneling processes for electrons with up spin and those with down spin. The 
characteristic times for these two processes are unequal due to polarization: there is 
fast tunneling for
spin-up electrons but slow tunneling for spin-down electrons
through the system. The electron spin which tunnel with a lower rate
modulates the tunneling of the other spin-direction electron (so-called
dynamical spin-blockade).\cite{Cottet} For a large
value of polarization, it leads to an effective bunching of
tunneling events and, consequently, to the super-Poissonian shot
noise.

It is worth noting that our findings in the P-configuration seem to be in obvious 
conflict wiht the recent prediction in the sequential tunneling through multi-level 
systems, in which it was argued that a negative differential conductance (NDC) surely 
accompanied by the classical super-Poissonian noise.\cite{Wacker,Axel} Actually, in 
these multi-level systems NDC can occur when two levels, separated by nonzero energy 
difference
$\Delta\epsilon=\epsilon_{2}-\epsilon_{1}$, have different
coupling strength to the leads. The tunneling through the level which has a stronger 
coupling to the leads will be
suppressed by the Coulomb interaction once the level with weaker
coupling is occupied by electron. This reduces the total current, thus induces the NDC. 
However, this is not the situation in this paper. Here we assume small Zeeman splitting 
of the level ($\Delta\epsilon=g\mu_{B}B<k_{B}T$) and no energy difference between two 
spin levels.

Further increasing the bias-voltage above the Coulomb blockade region,
i.e, for $eV/2>\epsilon_{d}+U$, opens one more conducting channel
and removes spin-blockade. In this region, spin-up and spin-down
electrons are tunneling through the different channels and there
is no more competition between these two tunneling events. This
leads to a reduction of the current fluctuation and the Fano
factor becomes the same as in the paramagnetic case.

The situation is completely different in the AP-configuration
[Fig.~1(b) and (d)]. An increase of the polarization increases
tunneling rates $\Gamma_{L\uparrow}$ and $\Gamma_{R\downarrow}$
and decreases tunneling rates $\Gamma_{L\downarrow}$ and
$\Gamma_{R\uparrow}$. An electron with the spin-up, which has
tunneled from the left electrode into the QD, remains there for a
long time because the tunneling rate $\Gamma_{R\uparrow}$ is
reduced by the polarization. This decreases the spin-up current.
An increase of the polarization also decreases the spin-down
current because it reduces the probability for tunneling of the
spin-down electrons into the QD. This will decrease the total
current through the system in the Coulomb blockade region, and at the same time increase 
the shot noise. For large voltage, in the region $eV/2>\epsilon_{d}+U$, both
conducting channels become available which results in reduction of
the noise comparing with the Coulomb blockade region. In this case
the Fano factor does not go to the paramagnetic value because the
asymmetry in the tunneling rates are still presented.

\begin{figure}[htb]
\includegraphics[height=3.in,width=3.5in]{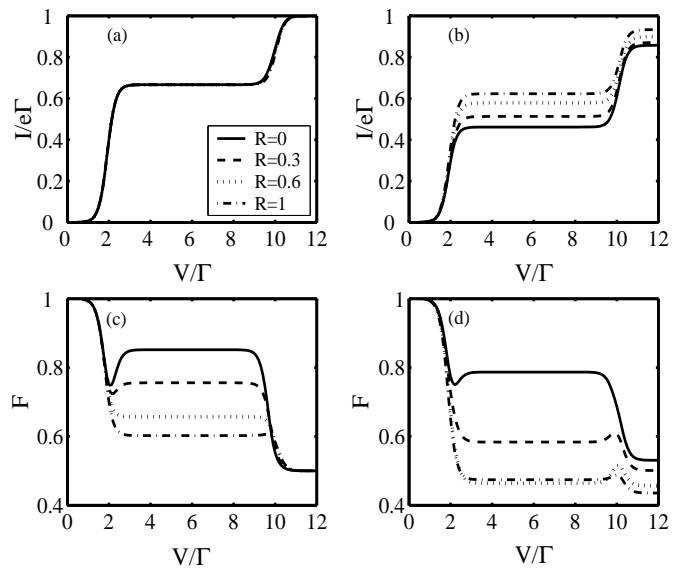}
\caption{Current (a, b) and Fano factor (c, d) in the P-configuration (a, c) and 
AP-configuration (b, d) calculated for
polarization $p=0.5$ and different spin-flip processes. Other
parameters are the same as in Fig.~\ref{FIG.1}.}\label{FIG.2}
\end{figure}

The dependence of Fano factor on the spin-flip scattering with given polarizations 
$p=0.5$ and $0.9$ can be found 
from Figs.~2 and 3, respectively. Introducing spin-flip scattering usually results in 
suppression of the zero-frequency current fluctuation
and the Fano factor. This behavior can be easily understood by the following 
consideration. The spin-flip scattering opens actually one new path for electrons to 
tunnel out of the QD: a spin-up(down) electron tunneling into the QD can now experience 
spin-related interaction insider the dot and change its spin-polarization direction and 
then exit from the QD as a spin-down(up) electron. As a result, an electron which is 
forced to spend more time in QD
due to polarized leads (for example, spin-down electron in the P-configuration and 
spin-up electron
in the AP-configuration) now has greater probability of leaving the dot. In other word, 
the spin-flip scattering plays a compensating role in tunneling in contrast to that of 
the lead polarizations and consequently makes an opposite contribution to the current 
fluctuations. However, this is not true for the P-configuration
when double occupation is allowed inside the QD (for
$eV/2>\epsilon_{d}+U$). In this region the spin-flip scattering does not have
any effect on the Fano factor [Figs.~2(c) and 3(c)]. Spin-up and
spin-down electrons are passing through separate channels without
changing their spins.

\begin{figure}[htb]
\includegraphics[height=3.in,width=3.5in]{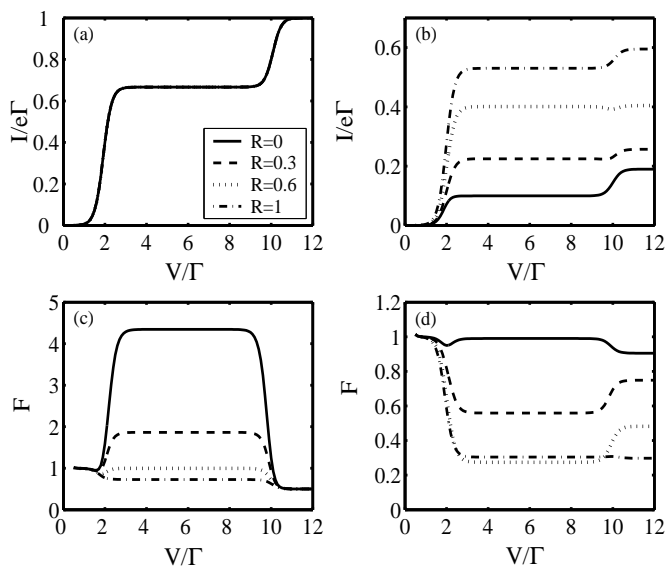}
\caption{The same as Fig.~\ref{FIG.2} except for
$p=0.9$.}\label{FIG.3}
\end{figure}

Fig.~4 shows the Fano factor vs. polarization. In the P-configuration, the Fano factor 
increases with polarization. In AP-configuration, for small spin-flip scattering, Fano 
factor increases with polarization but for larger $R$ it starts to
decrease. This is caused by competition of two effects: increase
in the Fano factor due to polarization and, decrease due to
spin-flip scattering.

\begin{figure}[htb]
\includegraphics[height=1.8in,width=3.5in]{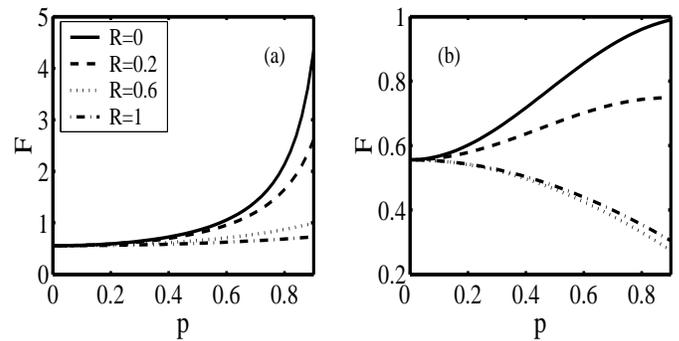}
\caption{Fano factor vs. polarization $p$ in
the P-configuration (a) and the AP-configuration (b) calculated
for different spin-flip processes $R$ in the Coulomb blockade
region, $V=4$. Other parameters are the same as in
Fig.~\ref{FIG.1}}\label{FIG.4}
\end{figure}

The frequency dependence of the Fano factor in the Coulomb
blockade region is given in Fig.~5. For weaker spin-flip scattering, increasing 
scattering leads to reduction in the Fano factor around
zero-frequency, in both configurations [Fig.~5(a) and (c)]. While the larger values of 
the spin-flip scattering cause different low-frequency behaviors for different leads 
polarizations [Fig.~5(b) and (d)]. In addition, the strong spin-flip scattering also 
generates an unambiguous peak in the P-configuration and a hump structure in the 
AP-configuration approximately located at a nonzero value of frequency $\omega=2R$ (the 
Rabi frequency), which characterizes the resonant oscillation between two spin states 
inside the QD. We provide a simple qualitative explanation for this behavior as 
following: When spin-polarized electrons are injected into the QD from the left lead, 
the spin Rabi oscillation always allows the electron to escape more easily from the dot 
to the right polarized lead, thus increasing the deviation of instant current from its 
average value, which induces enhancement or suppression of shot noise depending on the 
symmetry of states carrying current.\cite{GurvitzIEEE} 
In the P-configuration ($\Gamma_{R\uparrow}>\Gamma_{R\downarrow}$), the outgoing 
wavefunction of electrons preserve the symmetry with the dominated incoming electrons in 
the left lead [because we set $\Gamma_{L\uparrow}>\Gamma_{L\downarrow}$ for the 
polarization leads in this paper, see Eqs.~(\ref{pcon}) and (\ref{apcon})], which 
enhances the noise and thus causes a peak in the Fano factor [Fig.~5(a)]. Nevertheless, 
the situation is complicated for the AP-configuration 
($\Gamma_{R\uparrow}<\Gamma_{R\downarrow}$). The anti-symmetry generates a hump 
structure and finally a dip in the shot noise spectrum for high spin-flip scattering 
rate [Fig.~5(c)]. Moreover, in the high frequency region, the Fano factor goes to a 
constant $\frac{1}{2}$ for both configurations. 

\begin{figure}[htb]
\includegraphics[height=3in,width=3.5in]{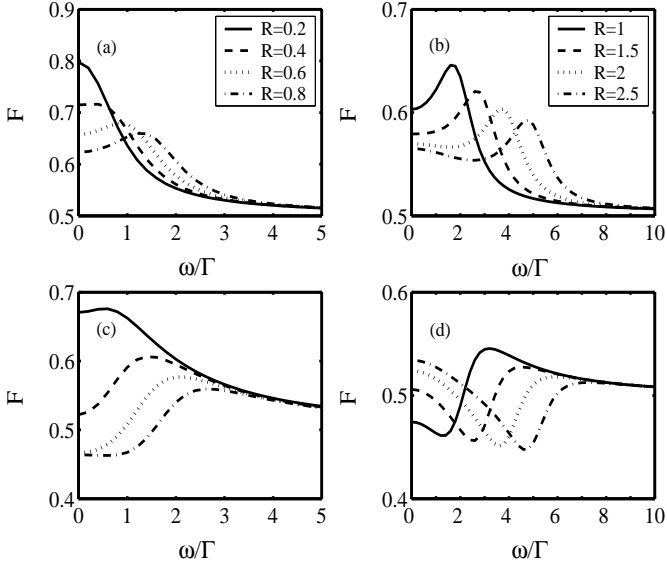}
\caption{Fano factor in the P-configuration (a, b) and AP-configuration (c, d) vs. 
frequency calculated in the Coulomb
blockade region for different spin-flip processes $R$. Other
parameters are: $\epsilon_{d}=1$, $U=4$, $V=3$, $p=0.5$, and
$T=0.1$.}\label{FIG.5}
\end{figure}

\section{Shot noise in COUPLED QUANTUM DOTS}

\subsection{Model and quantum rate equations}

In this section we consider a resonant tunneling through a CQD with weak coupling 
between the QDs and the leads. We assume that electron hopping $t$ between the two QDs 
is also weak. In order to simplify our derivation, we consider here the infinite 
intradot Coulomb repulsion $U'$ and a finite interdot Coulomb interaction $U$, which 
excludes the state with two electrons in the same QD but two electrons can occupy 
different QDs.
We are again interested in the regime $kT\ll U$, $\Delta$, and
$eV\ll\Delta$, so that only one level in each dot contribute to
the transport.

The tunneling Hamiltonian for the CQD is\cite{Dong} 
\bn 
H &=& \sum_{\eta, k, \sigma}\epsilon _{\eta k\sigma}
c_{\eta k\sigma }^{\dagger }c_{\eta k\sigma }^{\pdag}+ \epsilon_{1} \sum_{\sigma} c_{1 
\sigma }^{\dagger }c_{1 \sigma }^{\pdag}+ \epsilon_{2} \sum_{\sigma} c_{2 \sigma 
}^{\dagger }c_{2 \sigma }^{\pdag} \cr
&& + t\sum_{\sigma} (c_{1 \sigma}^{\dagger} c_{2\sigma}^{\pdag} + c_{2 \sigma}^{\dagger} 
c_{1\sigma}^{\pdag}) + U'n_{1 \uparrow }n_{1 \downarrow }+ U'n_{2
\uparrow }n_{2 \downarrow } \cr
&& + U \sum_{\sigma, \sigma'} n_{1 \sigma}n_{2 \sigma'}  +\sum_{k, \sigma} (V_{L \sigma} 
c_{L k\sigma }^{\dagger }c_{1 \sigma}^{\pdag} +{\rm {H.c.}}) \cr && + \sum_{k, \sigma} 
(V_{R \sigma} c_{R k\sigma }^{\dagger }c_{2 \sigma}^{\pdag} +{\rm
{H.c.}}), 
\label{hamiltonian3} 
\en
where $c_{1(2)\sigma}^{\dagger}$, $c_{1(2)\sigma}$ are creation and annihilation 
operators for a spin-$\sigma$ electron in the first (second) QD, respectively. 
$\epsilon_{i}$ ($i=1,2$) is the bare energy level of electrons in
the $i$th QD. The other notations are the same as in the SQD case.

Under the assumption of weak coupling between the QDs and the
leads, and with the application of the wide band limit in the two leads,
electronic transport through this system can be described by the
master equation.\cite{Dong} Here, in order to simplify the
analysis, we only consider spin independent tunneling processes
and we take the bare mismatch between the two bare levels to be
zero ($\epsilon_{1}=\epsilon_{2}=\epsilon_{d}$). The desired quantum rate
equations can be readily obtained as:
\begin{subequations}
\label{rateqCQD1} 
\bn
\dot{\rho}_{00}&=& \Gamma_{L}^{-} \rho_{11} + \Gamma_{R }^{-} \rho_{22} - 
(\Gamma_{L}^{+} + \Gamma_{R}^{+}) \rho_{00}, \cr
&& \label{rc01} \\
\dot{\rho}_{11}&=& \Gamma_{L}^{+} \rho_{00} + \widetilde{\Gamma}_{R}^{-} \rho_{dd} - 
\Gamma_{L}^{-} \rho_{11} \cr
&& - \widetilde{\Gamma}_{R}^{+} \rho_{11} + it( \rho_{12}-\rho_{21}), \label{rc11} \\
\dot{\rho}_{22}&=& \Gamma_{R}^{+} \rho_{00} + \widetilde{\Gamma}_{L}^{-} \rho_{dd}- 
\Gamma_{R}^{-} \rho_{22} \cr
&& - \widetilde{\Gamma}_{L}^{+} \rho_{22} - it( \rho_{12}-\rho_{21}), \label{rc21} \\
\dot{\rho}_{12}&=& +it (\rho_{11}- \rho_{22}) \cr
&& - \frac{1}{2} [\Gamma_{L}^{-} + \Gamma_{R}^{-} + \sum_{\eta} 
\widetilde{\Gamma}_{\eta}^{+}] \rho_{12}, \label{rc31} \\ 
\dot{\rho}_{dd}& =& \widetilde{\Gamma}_{R}^{+} \rho_{11} + \widetilde{\Gamma}_{L}^{+} 
\rho_{22} - (\widetilde{\Gamma}_{L}^{-} + \widetilde{\Gamma}_{R}^{-}) \rho_{dd}, 
\label{rc41} 
\en
\end{subequations}
with $\rho_{00} + \rho_{11} + \rho_{22} + \rho_{dd} = 1$.
Here, $\rho_{00}$ denotes the occupation probability that central region is empty, 
$\rho_{ii}$  ($i=1,2$) means that the $i$th QD is singly occupied by an electron, and 
$\rho_{dd}$ stands for the double occupation of the central region (each one of the QDs 
is occupied by one electron). The
non-diagonal elements $\rho_{ij}$ describe the superposition of
the two levels in different QDs. The tunneling rates $\Gamma_{\eta}^{\pm}= \Gamma_{\eta} 
f_{\eta}^{\pm}(\epsilon_{d})$ and $\widetilde{\Gamma}_{\eta }^{\pm}= \Gamma_{\eta} 
f_{\eta}^{\pm}(\epsilon_{d}+U)$ have the similar prescriptions as in the above section.

The electric current $I_{L}$ flowing from the lead $L$ to the QD
can be calculated as:
\bn 
I_{L}/e=\widetilde{\Gamma}_{L}^{-}
\rho_{dd} + \Gamma_{L}^{-}
\rho_{11} - \widetilde{\Gamma}_{L}^{+} \rho_{22} - \Gamma_{L}^{+} \rho_{00}.
\label{iiil}
\en
Similarly, for the
current flowing from the QD to the $R$ lead we have 
\bn 
I_{R}/e=-( \widetilde{\Gamma}_{R}^{-} \rho_{dd} + \Gamma_{R}^{-}
\rho_{22} - \widetilde{\Gamma}_{R}^{+} \rho_{11} - \Gamma_{R}^{+} \rho_{00}).
\label{iv}
\en
Again, the sign of the
current is chosen to be positive when the direction of the current
is from the left to the right.
It is easy to prove that, in stationary condition, the current conservation is fulfilled 
$I_{L}=I_{R}$.

\subsection{Discussions and calculations of shot noise}

In order to calculate the noise power spectrum in CQD, we can simply employ the same
procedure that we described in Sec.~II. First of all, the simplified quantum rate 
equations Eq.~(\ref{rateqCQD1}) can
be rewritten in the matrix form Eq.~(\ref{rate}), in which
${\rho}(t)=(\rho_{00}, \rho_{11}, \rho_{22}, \rho_{dd}, \rho_{12},
\rho_{21})^{T}$ and the matrix \textbf{M} can be read from
Eq.~(\ref{rateqCQD1}). Next, the current operators can be obtained from
Eqs.~(\ref{iiil}) and (\ref{iv}) as follows:
$\tilde{\Gamma}_{L}^{-}$ ($\tilde{\Gamma}_{R}^{-}$) describes a
tunneling of an electron from the first (second) QD to the
$L$$(R)$-lead and it can be considered as a transition rate from
$\rho_{dd}$ to $\rho_{22}$ ($\rho_{11}$). Similarly,
$\Gamma_{L}^{-}$ ($\Gamma_{R}^{-}$) is a transition rate from
$\rho_{11}$ ($\rho_{22}$) to $\rho_{00}$,
$\tilde{\Gamma}_{L}^{+}$ ($\tilde{\Gamma}_{R}^{+}$) describes
transition from $\rho_{22}$ ($\rho_{11}$) to $\rho_{dd}$, and
$\Gamma_{L}^{+}$ ($\Gamma_{R}^{+}$) is a transition rate from
$\rho_{00}$ to $\rho_{11}$ ($\rho_{22}$). In the basis
$(\rho_{00}, \rho_{11}, \rho_{22}, \rho_{dd}, \rho_{12},
\rho_{21})^{T}$, the left and right current operators are 
\bn
\hat{\Gamma}_{L}=\left(
\begin{array}{cccccc}
  0 & \Gamma_{L}^{-} & 0 & 0 & 0 & 0 \\
  -\Gamma_{L}^{+} & 0 & 0 & 0 & 0 & 0 \\
  0 & 0 & 0 & 0 & 0 & \tilde{\Gamma}_{L}^{-} \\
  0 & 0 & 0 & 0 & 0 & 0 \\
  0 & 0 & 0 & 0 & 0 & 0 \\
  0 & 0 & -\tilde{\Gamma}_{L}^{+} & 0 & 0 & 0 \\
\end{array}
\right),
\en 
and 
\bn
\hat{\Gamma}_{R}=\left(
\begin{array}{cccccc}
  0 & 0 & -\Gamma_{R}^{-} & 0 & 0 & 0 \\
  0 & 0 & 0 & 0 & 0 & -\tilde{\Gamma}_{R}^{-} \\
  \Gamma_{R}^{+} & 0 & 0 & 0 & 0 & 0 \\
  0 & 0 & 0 & 0 & 0 & 0 \\
  0 & 0 & 0 & 0 & 0 & 0 \\
  0 & \tilde{\Gamma}_{R}^{+} & 0 & 0 & 0 & 0 \\
\end{array}
\right),
\en
respectively. Finally, we can make corresponding substitution in 
Eqs.~(\ref{steadystate}), (\ref{noise}), and (\ref{Schottky}) to compute the quantum 
shot noise in CQD and use Eq.~(\ref{Fano factor}) to calculate the Fano factor.

In the following discussions, we set parameters of the CQD under consideration as:
$\epsilon_{d}=1$, Coulomb interaction $U=4$, and temperature $T=0.1$. And we also apply 
the bias voltage $V$ between two leads symmetrically, $eV/2=\mu_{L}=-\mu_{R}$. The zero 
of energy is chosen to be the Fermi level of the leads in the equilibrium and the energy 
unit is the one of the tunneling constants. 

First, we analyze the zero-frequency shot noise in three limits: (i) zero voltage limit; 
(ii) the Coulomb blockade region; and (iii) high voltage limit. 
When the bias voltage is below the resonance, i.e., $eV/2<\epsilon_{d}$, the
transport through the system is not energetically allowed and the
noise is Poissonian ($F=1$). As the bias voltage is increasing, there occur 
consecutively two plateaus in the current-voltage characteristic, separated by the 
thermally broadened step, corresponding to the case of $eV/2>\epsilon_{d}$ (the Coulomb 
blockade region) and of $eV/2>\epsilon_{d}+U$ (high voltage condition), respectively. On 
the first plateau, where only one
electron can be allowed inside the system, the current $I_1$ and the Fano
factor $F_1$ become:
\bq
I_{1}=\frac{4t^{2}\Gamma_{R}\Gamma_{L}}{\Gamma_{L}\Gamma_{R}^{2}+4t^2\left(2\Gamma_{L} + 
\Gamma_{R}\right)}, \label{I1}
\eq 
\bq
F_{1}=1-8\Gamma_{L}\Gamma_{R}t^{2}\frac{3\Gamma_{L}\Gamma_{R}+\Gamma_{R}^{2}+8t^{2}} 
{\left[4t^{2}\left(2\Gamma_{L}+\Gamma_{R}\right)+\Gamma_{L}\Gamma_{R}^{2}\right]^{2}}. 
\label{F1}
\eq
While in the second plateau, the CQD is in doubly occupied state, and the current $I_2$ 
and the Fano factor $F_2$ are: 
\bq
I_{2}=\frac{4t^{2}\Gamma_{R}\Gamma_{L}}{\left(\Gamma_{L}+\Gamma_{R}\right) 
\left(\Gamma_{L} \Gamma_{R}+4t^{2}\right)}, \label{I2}
\eq 
\bq
F_{2}=1-8\Gamma_{L}\Gamma_{R}t^{2}\frac{3\Gamma_{L}\Gamma_{R}+\Gamma_{R}^{2}+\Gamma_{L}^
{2}+4t^{2}}{\left(\Gamma_{L}+\Gamma_{R}\right)^{2}\left(\Gamma_{L}\Gamma_{R}+4t^{2} 
\right)^{2}}.\label{F2}
\eq
These results are in agreement with the previous derivation from the Laplace transform  
in Ref.~[\onlinecite{Brahim}].

We proceed with numerical calculations. In Fig.~6, we plot the currents and 
zero-frequency Fano factors in the CQDs as a function of bias voltage for various 
hopping rates between dots $t/\Gamma=0.2$, $0.5$, $0.7$, and $1.0$.  
We find that, depending on the relation between the hopping $t$ and the coupling
constants $\Gamma_{R(L)}$, the CQD manifests three distinct physical scenarios in its 
tunneling and low-frequency fluctuation properties: (i) In the case of hopping 
$2t>\sqrt{\Gamma_{L}\Gamma_{R}}$ ($t>1/2$ in Fig.~7), current increases in the second 
step ($I_{2}>I_{1}$) and the system exhibits positive differential conductance (PDC). 
Correspondingly, the Fano factor decreases ($F_2<F_1$) with increasing bias-voltage; 
(ii) If the hopping $2t=\sqrt{\Gamma_{L}\Gamma_{R}}$ ($t=1/2$ here), two currents and 
shot noise are equal to each other, respectively, at the entire region; (iii) While for 
the hopping $2t<\sqrt{\Gamma_{L}\Gamma_{R}}$ ($t<1/2$ here), current declines in the 
second step and consequently negative differential conductance (NDC) appears, but the 
Fano factor raises. Moreover, we find that the Fano factor is always below $1$ in all 
three regimes, indicating that the quantum shot noise is perpetually smaller than the 
classical value (sub-Poissonian). Once again, we obtain results differing from those by 
the recent papers Refs.~[\onlinecite{Wacker,Axel}]: Our shot noise is enhanced but does 
not reach super-Poissonian value in the NDC regime.

\begin{figure}[htb]
\includegraphics[height=3in,width=2.5in]{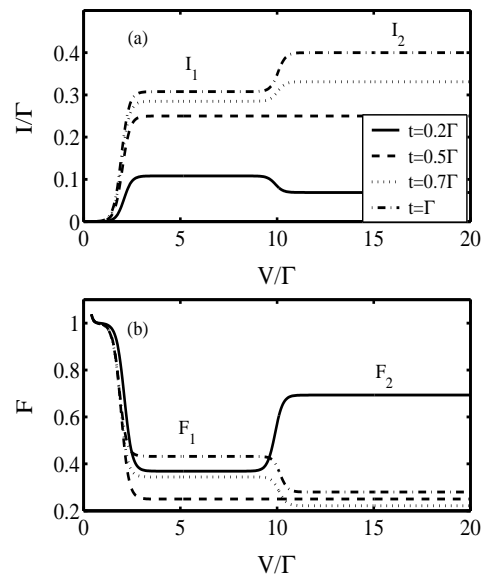}
 \caption{The calculated current (a) and Fano factor (b) vs. bias voltage
in the CQDs with different hoppings $t$ between two QDs.
Other parameters are: $\Gamma_{L}=\Gamma_{R}=\Gamma$,
$\epsilon_{d}=1$, $U=4$, and $T=0.1$.}\label{FIG.6}
\end{figure}

Fig.~7 and 8 shows the current and the Fano factor as a function
of the tunneling constants $\Gamma_{R}$ and $\Gamma_{L}$ in the
current plateaus regions. For small hopping $t<\Gamma_{L}$ the currents
$I_{1}$ and $I_{2}$ show non-monotonic behavior with increasing
$\Gamma_{R}$ [Fig.~7(a)]: They increase for small couplings and,
decrease when couplings are larger. With increasing $\Gamma_{L}$
[Fig.~7(b)], $I_{1}$ increases for small couplings and gradually saturates. On the 
contrary, $I_{2}$ shows an analogous
non-monotonic behavior as in Fig.~7(a). The NDC appears whenever $I_{2}$
becomes smaller than $I_{1}$ and it is most pronounced for
larger coupling to the left lead $\Gamma_{L}\gg \Gamma_{R}$. In small hopping limit 
($t\ll \Gamma_{L(R)}$),
Eqs.~(\ref{I1}) and (\ref{I2}) can be simplified, respectively, to:
\bn
I_{1}=\frac{4t^2}{\Gamma_{R}}=\frac{4t^2}{\Gamma_{d_{1}}}\label{II1},
\en
and
\bn
I_{2}=\frac{4t^2}{\Gamma_{R}+\Gamma_{L}}=\frac{4t^2}{\Gamma_{d_{2}}}\label{II2},
\en
where $\Gamma_{d_{1}}=\Gamma_{R}$ and
$\Gamma_{d_{2}}=\Gamma_{R}+\Gamma_{L}$ are the decoherence rates
due to the interaction with the leads. Actually, if we simplify 
Eq.~(\ref{rc31}) in the regions of the first and second current plateaus, we obtain:
\bq
\dot{\rho}_{12}=it(\rho_{11}-\rho_{22})-\frac{1}{2}\Gamma_{d_1} \rho_{12},
\label{qq1}
\eq
and
\bq
\dot{\rho}_{12}= it (\rho_{11}-\rho_{22}) - \frac{1}{2}\Gamma_{d_2} \rho_{12}, 
\label{qq2}
\eq
respectively. It is clear that the two rates  
indeed characterize the decay of coherent superposition.
Conservation of current allows us to express the current
through the system as $I=it(\rho_{12}-\rho_{21})$. From Eqs.~(\ref{qq1}) and (\ref{qq2}) 
one can see that for large decoherence
rates, $\Gamma_{d_{1(2)}}\gg t$, the probability of tunneling
between the two dots, $\rho_{12}$, becomes purely damped with time, implying the 
destructive quantum interference between the coherent
superposition of two quantum dot states and the electronic state in the continuum (right
lead) in the process of tunneling,\cite{Gur} which induces  
localization of the electron in the first dot (QDs form ionic-like bonds), as well as 
the reduction of the two resonant currents $I_1$ and $I_2$.

\begin{figure}[htb]
\includegraphics[height=3in,width=3.5in]{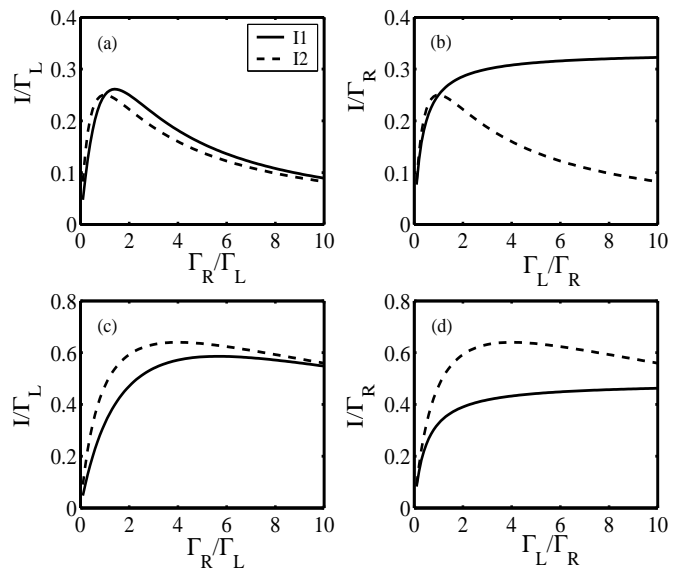}
 \caption{Current vs. bias-voltage
in the CQDs calculated for different coupling constants
$\Gamma_{R}$ (a, c) and $\Gamma_{L}$ (b, d). Other parameters are:
$U=4$, $\epsilon_{d}=1$, $T=0.1$, $t=0.5$ (a, b) and $t=2$ (c,
d).}\label{FIG.7}
\end{figure}

In the case of high voltage, the presence of one excess electron inside the dots
further destroys coherent superposition because there is no possibility of electron 
tunneling between the two
dots due to the infinite intradot Coulomb
interaction. 
The decoherence rate in this case is a sum of the two rates:
$\Gamma_{L}$, which describes the tunneling of the second electron
inside the system and $\Gamma_{R}$, which stands for the tunneling
of an electron to the right lead. The decrease of the current
$I_{2}$ is caused by both of these processes [Figs.~7(a) and (b)], therefore it is 
smaller that $I_1$ and NDC appears. In the NDC regime the noise increases
but it remains sub-Poissonian (Fig.~8).

\begin{figure}[htb]
\includegraphics[height=3in,width=3.5in]{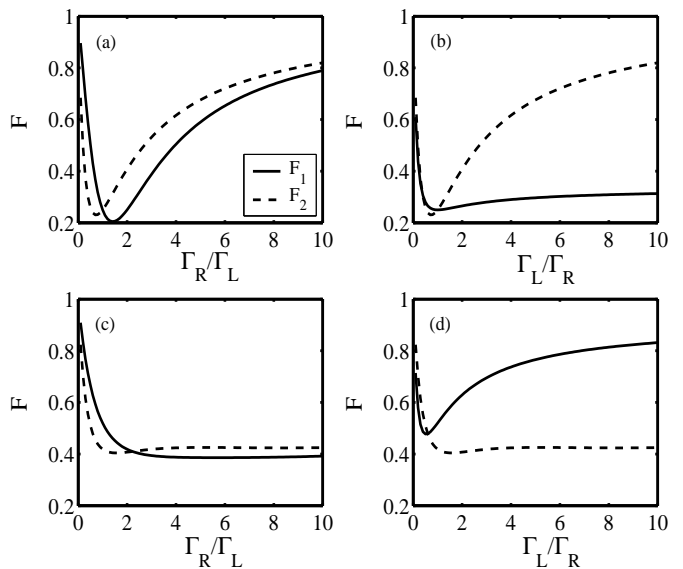}
 \caption{Fano factor vs. bias-voltage
in the CQDs calculated for different coupling constants
$\Gamma_{L}$ (a, c) and $\Gamma_{R}$ (b, d). Other parameters are
as in Fig.~\ref{FIG.7}}\label{FIG.8}
\end{figure}

In the large hopping limit ($t\gg\Gamma_{d_{1(2)}}$), an electron
can shuttle between the two dots many times in a phase-coherent way before it tunnels 
out into the right lead and thus
becomes delocalized (covalent-like bonds). In this limit, the
current and the Fano factor approach to the single QD values 
Eqs.~(\ref{current1.1})--(\ref{fano1.2}).

The Fano factor is plotted against normalized frequency in
Fig.~9. Similar noise properties are found for a coherently
coupled double well structure.\cite{Sun} In the Coulomb blockade
regime [Fig.~9], if an electron from the left lead is
injected into the first QD, no further electrons can enter in the
QD until this electron is removed. The time scale for this removal process
is determined by $t^{-1}$. Thus, in small hopping limit, the zero-frequency shot noise 
is reduced with increasing $t$ due to the interdot Coulomb blockade effect [Fig.~9(a)]. 
As the frequency increases, the electron inside the first QD has more opportunity to 
instantly tunnel into the second QD, which enhances the shot noise. 
While for the high values of hopping, the electron in the second QD can
either escape to the right lead, which takes place on a time scale
determined by $\Gamma_{R}^{-1}$, or, it can periodically return
to the first QD at a frequency $2t$ (the Rabi frequency). 
If $2t\gg\Gamma_{R}$, there is larger probability for the
electron in the second QD to tunnel back into the first QD than to
escape to the right lead, which prevent another electron from
entering the first QD. Thus at large values of $t$ noise suppression
occurs at $\omega=2t$, the dip in the Fano factor [Figs.~9(b)].

\begin{figure}[htb]
\includegraphics[height=1.5in,width=3.5in]{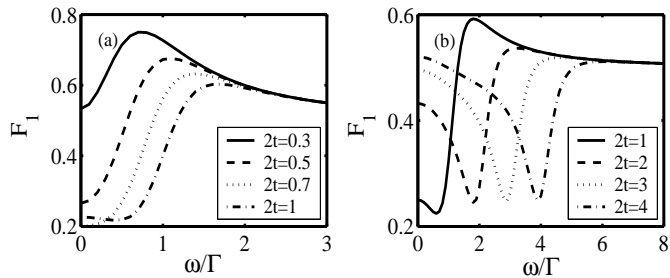}
 \caption{Fano factor vs. frequency calculated in the Coulomb
 blockade regime for different hoppings $t$ between the QDs.
 Other parameters are: $\Gamma_{R}=\Gamma_{L}=\Gamma$,
 $V=6$, $\epsilon_{d}=1$,
 $U=4$, and $T=0.1$.}\label{FIG.9}
\end{figure}

The influences of the coupling constants on the frequency-dependent noise characteristic 
is given in Fig.~10. When only one
electron is present inside the CQD system [Figs.~10(a) and (b)],
decreasing coupling to the right lead (decreasing decoherence rate)
leads to the noise enhancement in the low frequency region. On
the other hand, the decreasing decoherence rate results in the
transformation of the electronic states in the CQD from the ionic-like bond to the 
covalent-like bond and the formation of a dip in the shot noise spectrum at moderately 
high frequency $\omega=2t$. In the large voltage
region, when doubly occupancy is allowed, the decoherence rate is
given as a sum of $\Gamma_{L}$ and $\Gamma_{R}$ and both coupling
constants influence shot noise in the same way [Figs.~10(c) and
(d)].

\begin{figure}[htb]
\includegraphics[height=3in,width=3.5in]{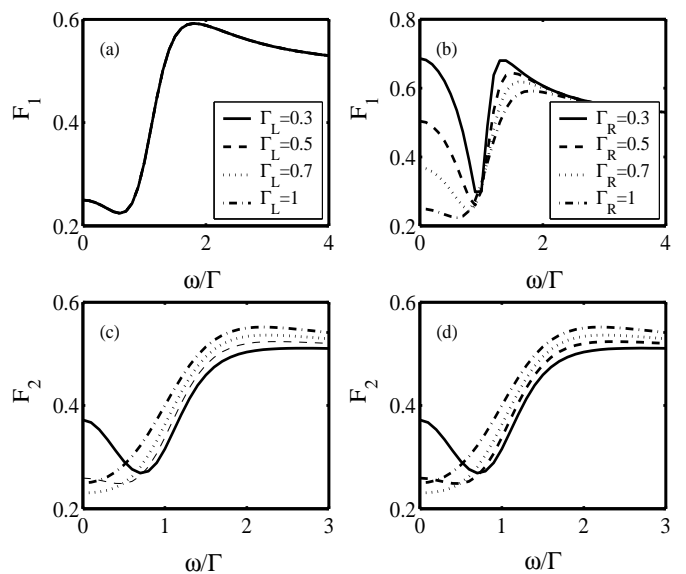}
 \caption{Fano factor vs. frequency calculated in the Coulomb
 blockade regime (a, b) and in the large voltage region
 (c, d) for different coupling constants $\Gamma_{L}$ and $\Gamma_{R}$
 between the QDs and the leads.
  Other parameters are: $t=0.5$, $\Gamma_{R}=\Gamma$ (a, c), $\Gamma_{L}=\Gamma$ (b, d)
  $V=6$ (a, b), $V=12$ (c, d), $\epsilon_{d}=1$,
  $U=4$, and $T=0.1$.}\label{FIG.10}
\end{figure}

\section{CONCLUSION}

In this paper we have presented theoretical investigations of the shot noise spectrum in 
resonant tunneling through a interacting quantum coupled system, in which quantum 
interference effects play an important role in its transport and fluctuation properties. 
For this purpose, we have modified the well-known generation-recombination approach, 
which has been established for over ten years to study the shot noise in single-electron 
tunneling devices based on the classical rate equations, to incorporate with the quantum 
version of rate equations. We have also developed a convenient numerical technique to 
compute the shot noise by applying the matrix spectral decomposition.  

As applications of our formalism, we have systematically analyzed the current shot 
noise, as functions of bias-voltage and frequency, through (i) a single QD connected to 
two ferromagnetic leads
with weak intradot spin-flip scattering, and (ii) coherently
coupled QDs. The influence of the Coulomb
interaction and coherent electron evolution on the Fano factor has
been investigated in detail. First we have given some analytic expressions for the 
zero-frequency Fano factors in two special cases: Coulomb blockade regime and double 
occupation region (high bias-voltage limit). It is shown that these results are in 
perfect agreement with previous analysis in the literature. 

In addition, we have performed numerical simulations for the two systems. For the single 
QD, we found that: 1) In the Coulomb
blockade regime, an increase of the polarization leads to an
enhancement of the zero-frequency current noise in both configurations, even a 
super-Poissonian value in the P-configuration due to dynamical spin-blockade, implying 
bunching effect in tunneling processes. For the
AP-configuration, this enhancement results from the asymmetry in the tunneling rates 
into and out of the QD ($\Gamma_{L\uparrow}>\Gamma_{R\uparrow}$ but
$\Gamma_{L\downarrow}<\Gamma_{R\downarrow}$) for each spin
separately; 2) The
spin-flip scattering compensates for polarization of the leads (removing the dynamical 
spin-blockade) and causes reduction of the current fluctuation and Fano factor; 3)
The frequency-dependent Fano factors clearly show a peak in the P-configuration but
a dip in the AP-configuration at the Rabi frequency $\omega=2R$, reflecting differing 
symmetries of states carrying current. For the CQD, we have predicted that depending on 
the relation between the hopping $t$ and the dot-leads couplings $\Gamma_{L(R)}$, there 
are three distinct physical scenarios in its tunneling and low-frequency fluctuation 
properties. More importantly, we found appearance of NDC in the current-voltage 
characteristic and corresponding enhancement of the Fano factor but still remaining 
sub-Poissonian in the case of $t<\sqrt{\Gamma_{L} \Gamma_{R}}$. Our sequent qualitative 
analysis claimed that this feature is a result of the enhanced decoherence rate induced 
by the presence of the second electron inside the system [Eqs.~(\ref{II1}) and 
(\ref{II2})]. Moreover, the destructive interference between two QDs states leads to a 
dip in the frequency-related Fano factor at the Rabi frequency $2t$, which is also 
controlled by the tunnel coupling $\Gamma_R$ to the right lead in the Coulomb blockade 
regime.      

Expectably, our results show that shot noise measurements can provide more useful 
information about quantum coherence of electron wavefunction than do conventional 
transport measurements.

\begin{acknowledgments} 

This work was supported by the DURINT Program administered by the US Army Research 
Office. The authors are grateful to B. Rosen for reading this manuscript. One of the 
authors, I. Djuric, acknowledges valuable discussions with L. Fedichkin, V. Puller, and 
Christopher Search.

\end{acknowledgments}

\end{document}